# Direct Observation of Channelised Supercurrents in a Kagome Superconductor

*Matthijs Rog[1]*, Tycho J. Blom[1]*, Reinier Q. Regter[1], Andrea Capa Salinas[2], Dalal Benali[3], Jinwon Lee[1,4], Daan B. Boltje[3], Mark H. Fischer[5], Titus Neupert[5], Stephen D. Wilson[2], Milan P. Allan[6,7,8], Kaveh Lahabi[1,3]*

1 Huygens-Kamerlingh Onnes Laboratory, Leiden University, 2300 RA Leiden, The Netherlands.
2 Materials Department, University of California Santa Barbara, Santa Barbara, California 93106, USA.
3 QuantaMap B.V., Robert Boyleweg 4, 2333 CG Leiden, The Netherlands.
4 Department of Quantum Nanoscience, Delft University of Technology, 2628 CJ Delft, The Netherlands
5 Department of Physics, University of Zürich, Winterthurerstrasse 190, 8057 Zürich, Switzerland
6 Faculty of Physics, Ludwig-Maximilians-University Munich, Munich 80799, Germany
7 Center for Nano Science (CeNS), Ludwig-Maximilians-University Munich, Munich 80799, Germany
8 Munich Center for Quantum Science and Technology (MCQST), Ludwig-Maximilians-University Munich, Munich 80799, Germany

*) These authors contributed equally.

## Abstract

Superconductors are many-body quantum states in which current flows without dissipation. Theory predicts that supercurrents follow a relatively simple spatial pattern in both conventional and unconventional superconductors. Recent studies into the $AV_3Sb_5$ (A = Cs, K, Rb) family of Kagome superconductors indicate that $CsV_3Sb_5$ has unconventional transport properties that cannot be accounted for with these simple theories, including reports of intrinsic Josephson junctions, higher order Cooper pairing and the zero field diode effect[1–5]. Attempts to interpret these findings have focused on the interplay of superconductivity with the unconventional charge density wave (CDW) order in these materials, with which superconductivity competes[6–8]. A current roadblock to understanding how these kagome superconductors give rise to their intriguing properties is the lack of spatially resolved information about transport. Here we show, using a recently developed superconducting quantum interference device (SQUID) microscope, that flakes of $CsV_3Sb_{5-x}Sn_x$ host a network of narrow supercurrent channels. These supercurrent channels emerge at the critical temperature and remain stable for all temperatures and currents. Their non-linear behaviour is consistent with a network of Josephson junctions linked by narrow supercurrent filaments, which naturally leads to the observed transport anomalies. Intriguingly, these observations are much weaker in undoped samples, which suggests links to the physics of charge density waves, disorder, and electronic correlations, all of which are greatly influenced by the doping strength. These results establish new frontiers for the local investigation of charge transport and competing orders in strongly correlated electron systems, and shine a new light on the anomalous transport properties of the $AV_3Sb_5$ kagome superconductors.

## Introduction

In quantum materials, non-trivial aspects of quantum mechanics such as quantum condensation or collective topological states can emerge naturally as part of the complex mechanics of the system. Quantum materials are intensely studied for their diverse technological applications, which include quantum computing and sensing. Still, although many have been identified, the high complexity of these materials, as well as their complex intertwining orders, means that we are far from a comprehensive understanding of these materials.

One heavily-studied family of quantum materials is the $AV_3Sb_5$ (A = K, Rb, Cs) family of non-magnetic kagome superconductors. This group of materials has attracted intense experimental and theoretical investigation, owing to its band structure that combines topology, flat bands, and Dirac cones, as well as its exotic charge density wave (CDW) state[9–12], which has led to proposals of loop current order[12–15]. Moreover, the appearance of superconductivity at lower temperatures brings up exciting questions about its interplay with the CDW[16–19]. Much of the important progress towards a comprehensive understanding of the superconducting state in these materials has been achieved through transport experiments on single-crystal flakes, which have yielded numerous unexplained observations, including the superconducting diode effect[2,4], intrinsic supercurrent interference[3–5], half-quantum flux states[20] and signatures of higher charge flux quanta[1].

However, a critical hurdle towards understanding the transport experiments in $AV_3Sb_5$ is the current dearth of local information about electrical transport. Generally, spatially resolved information is crucial to understand the underlying physics of quantum materials, because strong correlations and competing orders dominate the electronic landscape[21]. In addition, mesoscopic devices involving quantum materials may be influenced by geometric factors whose effects may be misinterpreted as representing bulk physics of the material absent local information. In contrast to the ubiquity of local spectroscopic and topographic characterisation, techniques that can probe nanoscale transport effectively in real devices are scarce[22]. The recent development of the tapping-mode SQUID-on-tip (TM-SOT) microscope offers a solution, as its combination of SQUID-on-tip imaging and tapping mode atomic force microscopy (AFM) makes it uniquely suited for current imaging in devices and flakes due to its ability to correlate nanoscale transport with the topography of the system[22].

In this study, we investigate supercurrents in flakes (thickness 60-140 nm) of the doped kagome superconductor $CsV_3Sb_{5-x}Sn_x$ (Fig. 1a) by local magnetic field imaging. Our results show that the supercurrents in these flakes are exclusively confined to narrow channels, while normal state currents are homogeneously distributed. Beyond transport currents, we observe that Meissner and vortex currents are confined to the same narrow regions. These current-carrying channels, which emerge at the onset of superconductivity (where the resistivity first drops), remain stable over a large range of currents and temperatures, and are much narrower than all relevant electromagnetic length scales. The results establish a firm connection between the previously-observed interference patterns in $CsV_3Sb_{5-x}Sn_x$ flakes and the channelised current distribution.

A crucial aspect of our findings is that the channelised supercurrent network only emerges in samples of light hole-doping, while the first hints of channelisation are present in an undoped flake. Hole-doping tunes both charge order and superconductivity[17,23,24], allowing for systematic shifting of features in the electronic landscape. Even light doping, achieved through Sn substitution of Sb, significantly alters the shape and species of the parent CDW[23,25,26] and supresses long-range charge order[17,27].

## Emergence of Channelised Supercurrents

To arrive at these results, we prepared two doped ($x$ = 0.03) flakes (devices D1 and D2) through mechanical exfoliation from the parent crystal shown in Fig. 1b. In the parent crystal, scanning tunnelling spectroscopy at various positions, shown in Extended Data Fig. 2, confirms bulk superconductivity consistent with previous reports. The structural and chemical homogeneity of our flake devices was confirmed in our earlier work[5]. The onset of the CDW phase and the superconducting transition of these devices are clearly visible in the resistance-temperature curve (Fig. 1d).

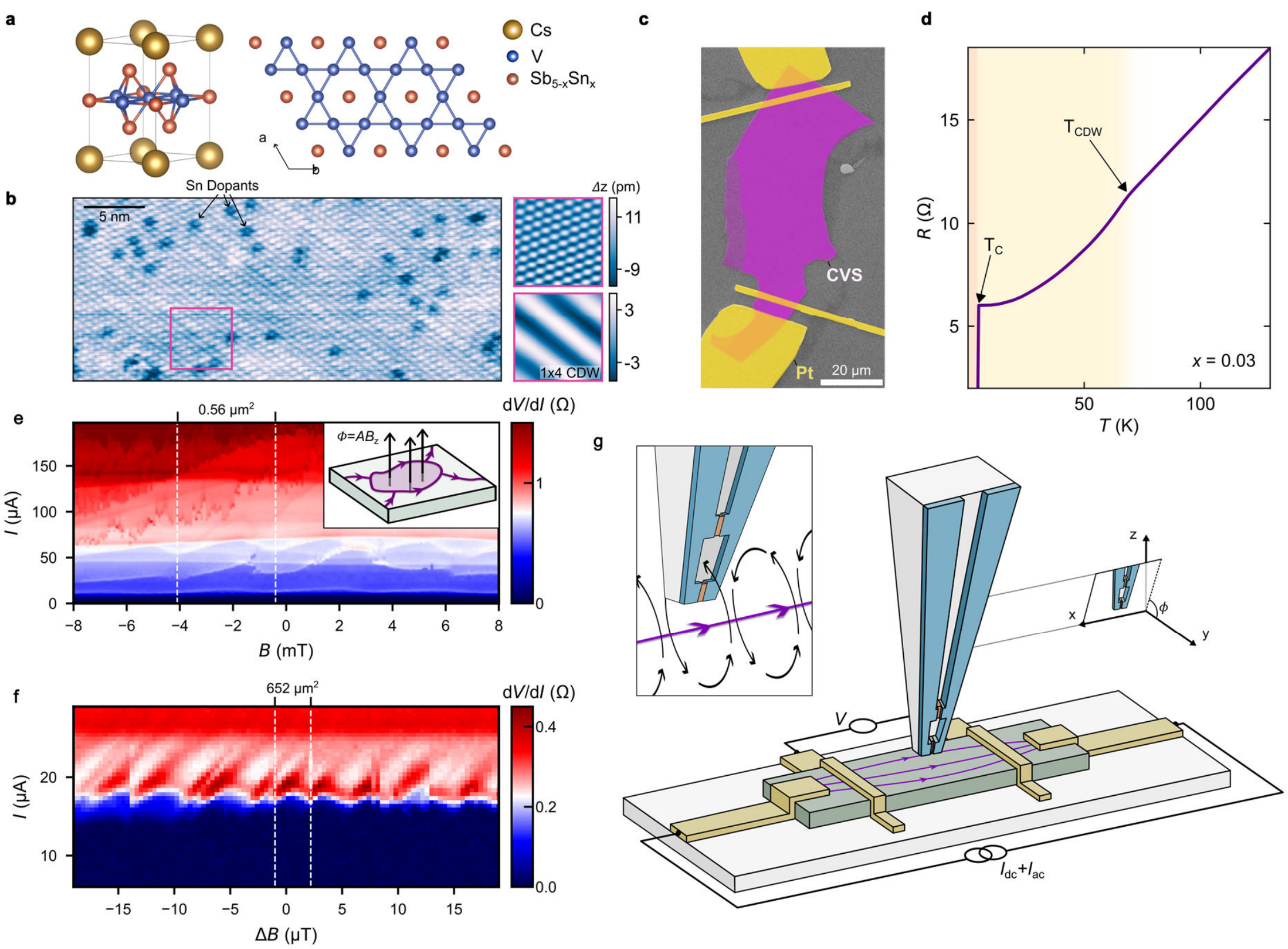

**Fig. 1 | Flakes of $CsV_3Sb_{5-x}Sn_x$. a,** Crystal structure of $CsV_3Sb_{5-x}Sn_x$. **b**, Atomic-resolution STM topograph on the Sb-termination of the $x$ = 0.03 parent crystal, showing the distribution of Sn dopants. The right images show the underlying Sb lattice and the 1x4 CDW structure inside the magenta rectangle, obtained through Fourier-space filtering. **c**, False coloured scanning electron image of the device D1 ($x$ = 0.03), contacted with Pt leads. **d**, Resistance-temperature measurement on device D1, emphasizing the transition to a CDW and then a superconducting state. **e**,**f**, Supercurrent interference patterns on device D1, showing that there are intrinsic oscillations over a wide range of periods. The inset depicts an illustration of supercurrent interference between distinct current paths. The intersections between these paths form loops through which the flux $\Phi$ drives the interference. **g**, Illustration of the tapping-mode SQUID-on-tip experiment on a flake. The inset shows that the SQUID locally maps current flow by capturing the in-plane component of the induced field $B_{ind}$.

In Fig. 1e,f, we show supercurrent interference patterns measured on flake D1 (Fig. 1c). Many oscillating critical currents can be distinguished, all with different periods spanning three orders of magnitude. This effect, for which various conflicting explanations have been proposed[2–5], is the first indication of an underlying inhomogeneous current flow. Intriguingly, the supercurrent appears to follow well-defined paths. The period of the oscillations could in principle correspond to a single flux quantum through a loop spanned by the interfering supercurrent paths, as illustrated in the inset. As the periods observed in the interference pattern vary by three orders of magnitude, this would require many interference loops of widely different sizes present in the flakes. We emphasize that these interference effects are not limited to some specific range of temperatures; the interference effects have been observed both close to $T_c$ and down to 50 mK, indicating that the effect stems from an intrinsic property of the superconducting ground state[4].

To locally map the flow of supercurrent in these flakes, we use TM-SOT microscopy. The experimental setup is illustrated in Fig. 1g. This technique uses a tilted SQUID-sensor ($\phi$ = 63°) which emphasizes in-plane magnetic field, making it uniquely suited for visualising small currents in flakes and nanodevices at high spatial resolution[22]. Moreover, the inclusion of tapping-mode AFM gives it the unique ability to map device topography alongside the magnetic measurements and correlate observed features to step edges and flake boundaries.

In Fig. 2, we visualize the current transport in device D1 through its superconducting transition. Fig. 2a shows its relatively broad transition, which contains multiple kinks and plateaus in the resistance. We probe the current distribution at three different temperatures: above $T_c$ (Fig. 2b), within the resistive transition (Fig. 2c), and below $T$ ~ 2K (Fig. 2d), which is the point at which the resistance reaches zero. We extract the current profile along a linecut (marked with arrows) using an iterative Biot-Savart model described in the Supplementary Information.

These measurements show the unexpected emergence of a channelised supercurrent network, which carries the supercurrent below the transition. Above $T_c$, the current transport is still completely homogeneous, demonstrating the metallic behaviour of the flake. As we cool down through the transition, supercurrent channels begin to emerge. While still in a resistive state, transport is a mixture of uniformly distributed ohmic currents and spatially confined supercurrents. Below $T_c$, the ohmic background vanishes and a network of channelised supercurrents dominates the transport.

Although the channels are distributed throughout the flake, they are more prevalent at, or near, the sample edges. We identify three main channels: one at the right edge, one at the left edge, and one at the step edge marked in red (Fig. 2a). However, not all channel locations correlate with topographic features. A number of thin filaments exist in the interior of the flake, while AFM shows that the top surface of the flake is free of any step edges or terraces (Extended Data Fig. 3). However, it cannot be excluded that step edges exist at the bottom surface at these locations. Importantly, the effect is robust to thermal cycling up to room temperature, with the current channels emerging at the same exact location upon cooling below the superconducting transition.

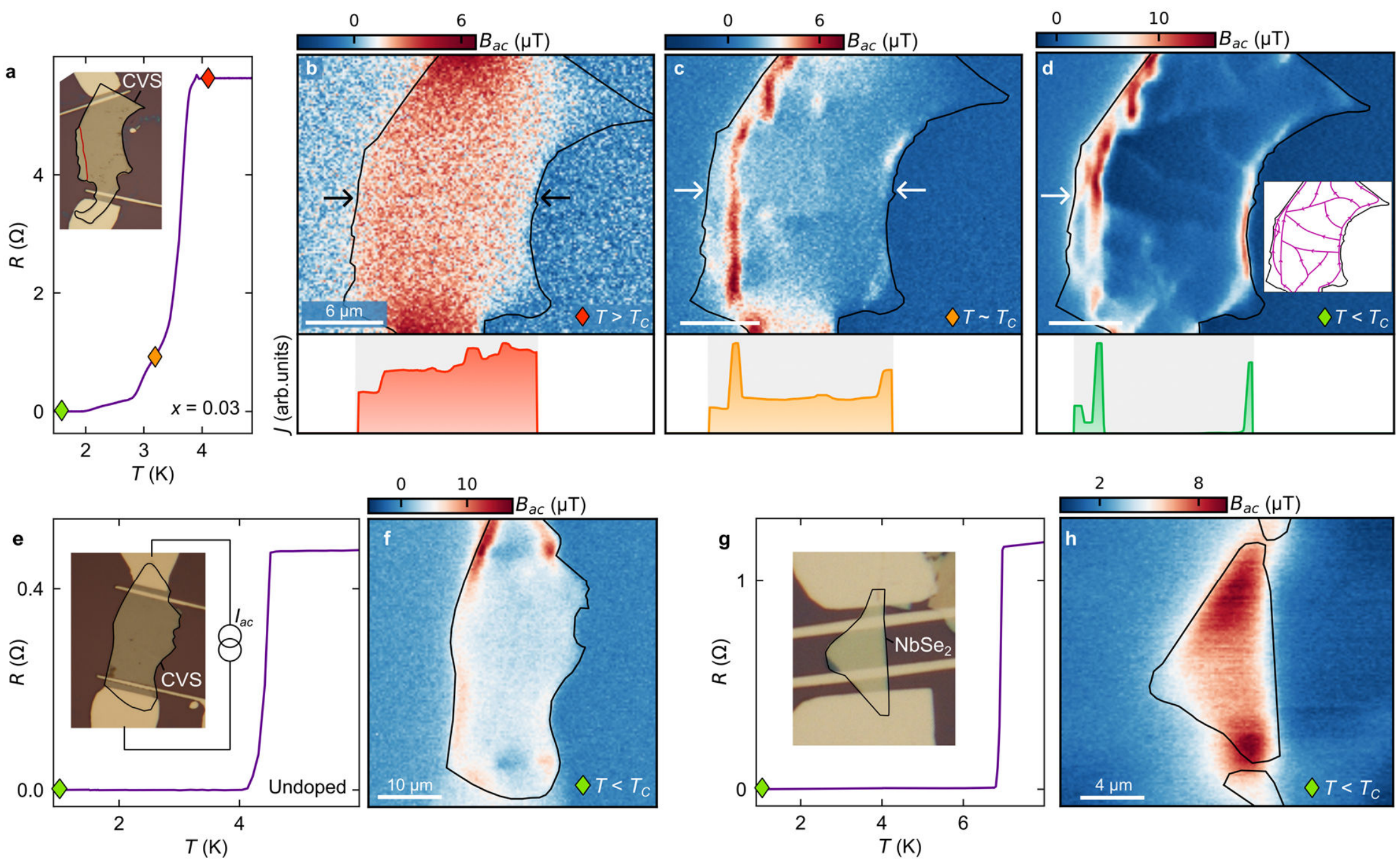

**Fig. 2 | Channelised supercurrents in $CsV_3Sb_{5-x}Sn_x$. a**, Resistance-temperature characteristic of device D1 through its superconducting transition. Before it hits the zero resistance state, it discontinuously passes through various features. **b-d**, TM-SOT images of the magnetic field $B_{ac}$ generated by an AC current $I_{ac} = 85$ μA applied to the flake, for the three temperatures marked in **a**. Channelised supercurrent paths emerge on the transition and, at the lowest temperature, span the entire flake. Below the images, current patterns along a line cut, marked by the arrows, are shown. The current patterns are normalized to their peak value. The inset of **d** shows the current paths identified from the SOT image. **e**, Resistance-temperature characteristic of device D3 (undoped). The transition is much sharper than that in **a**. **f**, TM-SOT image of device D4 taken deep in the superconducting state. Near the platinum contacts, supercurrent strongly crowds the flake edge. In the bulk, however, the effects of inhomogeneity are far weaker than for light doping. $I_{ac} = 210$ μA **g**,**h**, the same study as in **e**,**f**, performed on a $NbSe_2$ flake. $I_{ac} = 40$ μA

Extended Data Fig. 1 shows the same study performed on a second doped device (D2). This flake is smaller than D1, with a simpler geometry. Additionally, it has a small Pt deposit on its right edge, as an artifact of the contacting process. Here, too, a homogeneous current flow develops into a channelised, confined supercurrent flow through the superconducting transition. However, the network that appears in this device is much simpler, and we identify only one main channel at the right edge. We point out that the left edge, which attracts no channel, is densely terraced (Extended Data Fig. 3). The presence of the Pt deposit has a pronounced effect on the current flow: the current briefly stops tracking the right edge, and instead flows around the metallic deposit. A similar effect is visible at the top voltage contact, where part of the current tracks the Pt lead instead of the flake edge. These findings suggest that the underlying generator of the channelised structure is affected by the sputtered metallic film. Moreover, we emphasize that in both devices D1 and D2, we observe no temperature dependence of the channel positions.

Comparing the supercurrent interference patterns (Fig. 1e,f, Extended Data Fig. 1b,c), we see a direct correlation between the oscillations and the channelised supercurrent distributions. The many different channels in device D1 correspond directly to its complex interference pattern.

Here, the smallest oscillation period (641 μm$^2$) matches the area of the flake. Our imaging reveals that this is no coincidence: the outer edges of the flake host supercurrent channels that lead to this two-channel interference pattern. In contrast, the interference pattern of device D2 is much simpler: for large field ranges (small areas), there are no clear signs of two-channel interference. At small field ranges (large areas), it shows a very small ($\Delta I_c/I_c \sim 1$ %) modulation of the critical current, again matching the total flake area (241 μm$^2$). This corresponds to the observed current distribution, where the right edge channel strongly dominates the transport, and the left channel carries far less current. In this case, the entire flake behaves as a SQUID, where the two channels are highly asymmetric in $I_c$ and inductance.

Given that there are conflicting reports about the presence of the interference effect in undoped flakes from previous transport experiments[2–5], we also perform imaging experiments on an undoped flake (D3, Fig. 2e,f). Here, we do not observe any strongly confined channelised network. Near the contacts, however, the supercurrent is strongly inhomogeneous, with little to no current flow in the interior. We can tentatively connect this observation with the effect of the Pt deposit in device D2, further suggesting that the sputtered metallic films affect the generator of the current inhomogeneity. The effect rapidly decays away from the voltage probes. Away from the contacts, there is still supercurrent accumulation at the edges, most prominently on the left edge, but these currents decay over longer distances, similar to the edge-accumulation expected from London theory (Extended Data Fig. 4). Still, comparing to our simulations, we find that the edge currents are more enhanced than can be expected from London theory alone, indicating that the current distribution in the undoped flake is already non-trivial. The lack of a channelised current distribution is fully consistent with the supercurrent interference pattern, which shows no clear signatures of two-channel interference (Extended Data Fig. 4).

As a control, we also probed a flake of 2H-$NbSe_2$, which is the prototypical material where superconductivity coexists with a CDW. Fig. 2g,h shows that such a flake exclusively hosts uniform currents. In Extended Data Fig. 5, we show this current distribution is fully consistent with London theory.

## Evolution of Supercurrent Distribution

In a superconducting thin film, current is not expected to be distributed fully homogeneously. Instead, the current density is larger at the edges and decays over the Pearl length $\Lambda(T) = 2\lambda^2/d$, where $\lambda(T)$ is the London penetration depth and $d < \lambda$ the film thickness[28]. In Fig. 3a we show that this natural non-uniformity cannot account for the observed effects, as the supercurrent channels we observe in the doped flake are significantly smaller than the Pearl length. We use $\lambda(0) = 460$ nm[29] and the temperature dependence measured by Grant *et al*[30]. Moreover, their width is independent of temperature, which makes it highly unlikely that they relate to one of the fundamental length scales of superconductivity, which all diverge at $T_c$.

We then investigate the effect of the injected current $I_{inj}$ on the distribution of supercurrents. From transport, we know that this flake exhibits a cascade of critical currents, shown in Fig. 3b. We emphasize that none of these critical currents correspond to the depairing current of a thin superconducting nanowire. At the depairing current, Cooper pairs break up and the material reverts to its normal, metallic state. In both our images (Extended Data Fig. 8) and the analysis here, we never observe the destruction of a supercurrent filament despite crossing many critical currents. We must therefore conclude these critical currents are not depairing currents. As we increase $I_{inj}$ across the many critical currents, we never reach the normal state of the flake with $R \sim 6\ \Omega$.

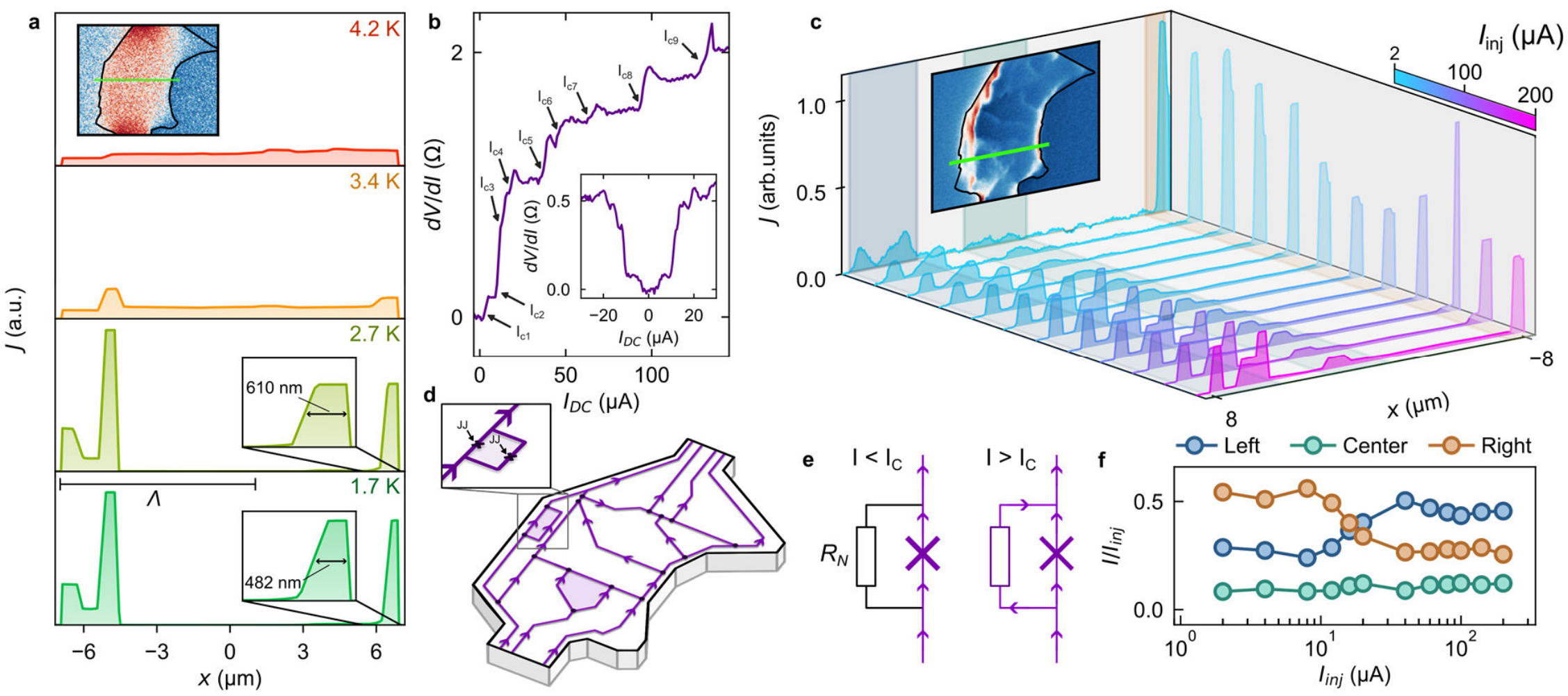

**Fig. 3 | Evolution of the channelised current distribution in device D1. a**, Current distribution at different temperatures through the transition along the line marked in green. All curves are normalized to their area. The channels appear immediately at the transition, are stable in temperature, and have a width smaller than the Pearl length Λ. **b**, Transport characteristics of device D1 at 1.7 K. The flake hosts a cascade of critical currents, marked by the black arrows. **c**, Evolution of the current distribution along the line marked in green, as the injected current is increased from 2 μA ($I < I_{c1}$) to 200 μA ($I > I_{c9}$). While currents redistribute and a small interior current appears, the channels never break up, shift, or change their width. **d**, Illustration of the resistively shunted junction (RSJ) model, which explains the emergence of a bulk current without destruction of superconductivity in the channels. Above the critical current, current is shared between a supercurrent inside the junction and a normal current through the parallel quasiparticle channel. **e**, Distribution of the current between the three major channels marked in **c**. The non-linear evolution of the current distribution indicates there is an underlying network of Josephson junctions.

The evolution of the normalized current distribution across a line-cut as we ramp up $I_{\text{inj}}$ (Fig. 3c) shows a redistribution of current across the channels and the flake interior. While the amount of current inside each channel changes, the size of the channels stays constant from below $I_{c1}$ up to 200 μA. The proportion of current through each channel changes non-monotonically, and at no point does a channel saturate. The fact that the channels do not disappear, regardless of how many critical currents are crossed, reaffirms that these critical currents are not related to the destruction of the superconducting condensate.

The cascade of critical currents, the non-monotonic evolution of the supercurrent distribution and the robustness of the channels to temperature and bias current are all consistent with a description of the critical currents as belonging to intrinsic Josephson junctions connected by nanowires, and not to a phase-coherent nanowire network alone. Fig. 3d depicts this interpretation: the channelised currents flow in a multiply connected network, which contains Josephson junctions along the paths. Parallel paths, as shown in the inset, form SQUID-loops that generate the observed interference effects. In this picture, the non-linear evolution of the current distribution emerges naturally. Unlike superconducting nanowires, Josephson junctions have a strongly non-linear current-phase relation which naturally results in a re-distribution of supercurrent as the applied current is increased. Moreover, in the Josephson junction picture, the junction always carries current, regardless of the magnitude of the injected current. We illustrate this in Fig. 3e with a simple toy model of a Josephson junction: the resistively shunted junction. As the critical current of the junction is reached and a voltage arises over the junction, a quasiparticle current flows parallel to the junction in addition to the supercurrent. Since quasiparticles can flow anywhere in these flakes, this implies that, as soon as the first critical current is passed, a homogeneously distributed bulk current $I_{\mathrm{bulk}}$ must flow alongside the persistent channels. This homogeneous background current is indeed present in Fig. 3c. From the current distributions, we infer the division of the current between the three dominant channels marked in Fig. 3c. Fig. 3e shows how this division non-monotonically evolves with $I_{\mathrm{inj}}$. While the right channel initially dominates the transport, it is overtaken by the two channels on the left around 15 μA. These conclusions shine a new light on the observation of Shapiro steps in this flake, which we show in Extended Data Fig. 6, and was reported previously [3–5].

## Magnetic Response

So far, we have only investigated the spatial distribution of externally applied transport currents. As an additional, independent probe of coherence, we also investigate the response of the flakes to an externally applied magnetic field. The magnetic response of a thin superconductor manifests as the partial screening of external fields over the Pearl length Λ, and the generation of Pearl vortices in the interior. We study the magnetic response inside device D1 by measuring the field gradient $dB/dz$ above the flake after field-cooling in a small field ($H \leq 4$ G). In Fig. 4a,b we observe a complex magnetic response, containing vortices as well as screening fields, that is completely confined to the same narrow regions where supercurrent flows. However, both Meissner currents and vortices only emerge on the left side of the flake, where the channels are wider. In both the interior and on the right edge, where the channels are narrow in comparison, no shielding effects are observed.

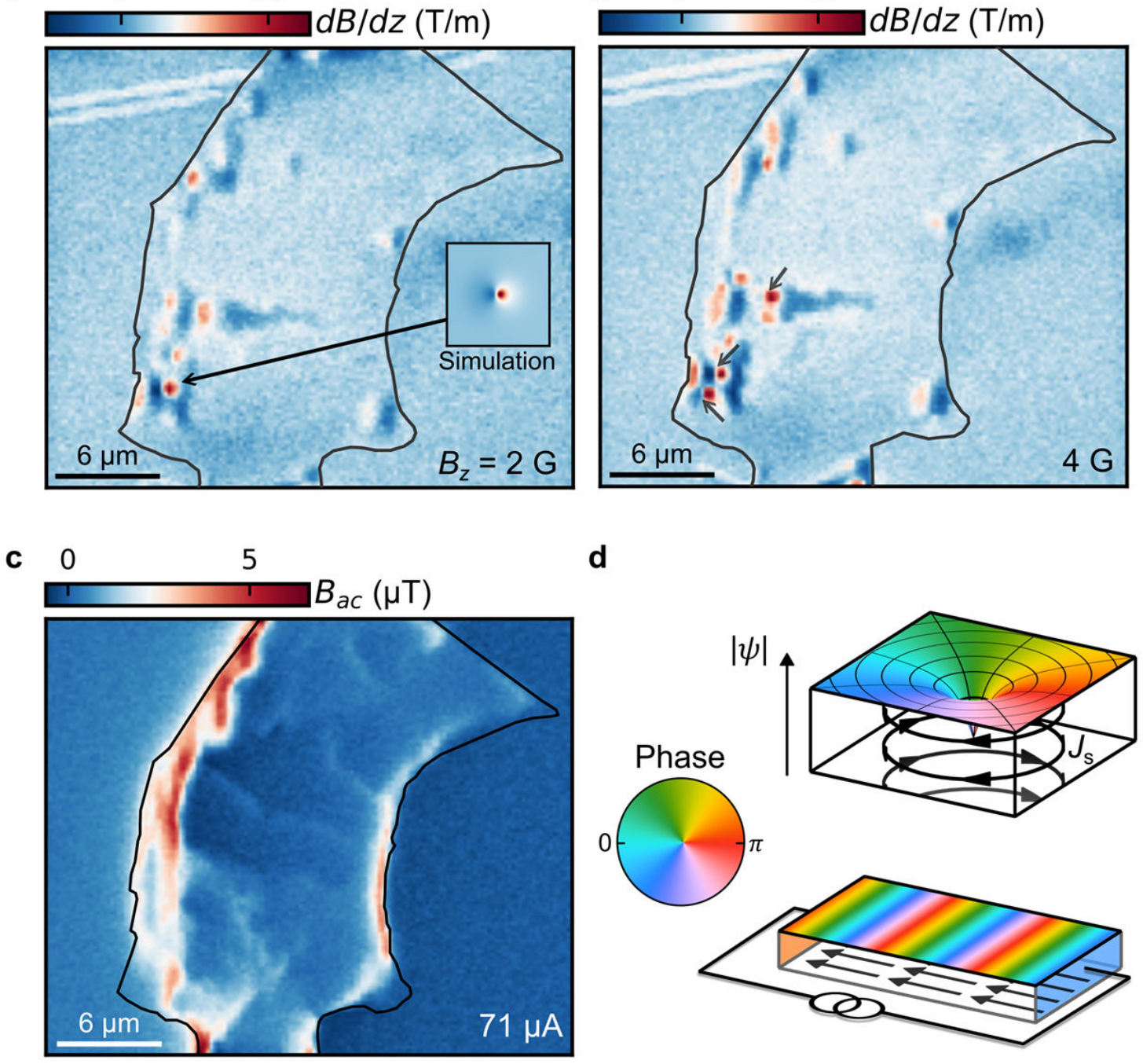

**Fig. 4 | Inhomogeneous Meissner screening and vortex generation. a**,**b**, TM-SOT images of the screening fields on device D1, taken in a small out-of-plane field, revealing both Meissner currents and vortices. The image, taken in gradiometric imaging mode, shows the out-of-plane gradients of the SQUID signal. A simulation of a vortex recorded by the in-plane (70°) TM-SOT probe is an inset in **a**. **c**, TM-SOT image of the channelised supercurrent flow inside the same field-of-view. Comparing the images, all vortices Meissner currents are confined to the widest current-carrying channels. **d**,

Illustrations of a Pearl vortex and a current-carrying superconducting wire, showing that both are driven by phase gradients. This implies that outside the current-carrying channels, there is no coherent condensate.

These observations show that all supercurrents in the flake, whether externally injected or intrinsically present, are confined to the channels. In a superconducting condensate, both supercurrents and vortices require gradients in the phase of the superconducting order parameter (Fig. 4d). The magnetic response is therefore a second, independent spatial probe of phase coherence. This demonstrates that a coherent condensate has only developed along these channels. In Extended Data Fig. 7 we show that the location of the channels is stable in fields up to at least 50 mT. While the distribution of supercurrent through the network changes, the channel locations remain fixed.

## Discussion

Our results show that all supercurrents in our $CsV_3Sb_{5-x}Sn_x$ (x = 0.03) flakes are carried in narrow channels, which display non-linearities consistent with the presence of Josephson junctions along their length. These results give a natural explanation for the anomalous transport data in flakes reported previously. Our observations are in stark contrast to the behaviour expected for a typical superconducting flake, as we verified in our study of a $NbSe_2$ device. We will now proceed to discuss the possible origin of the channelised supercurrents, the links between superconductivity and other local orders that exist in these materials, as well as important questions that our observations raise.

As shown, the supercurrent channels are stable with temperature. Once a channel is formed, it does not move or change size upon reduction of the temperature, despite the significant temperature dependences of the relevant superconducting length scales. This implies that the formation of the channels is strongly connected to an underlying order. Because the effect does not modulate with temperature, even when cycling to 300 K, this underlying order appears to be structural. As our undoped flakes do not exhibit the same degree of channelisation, the structural order underlying the effect appears to be connected to doping.

From our observations, we cannot with certainty identify the microscopic mechanism that leads to the supercurrent channels in Sn-doped flakes. Hints might come from the effects that Sn-doping has on the physics of $CsV_3Sb_5$. In addition to increasing the disorder, hole-doping has a substantial effect on the charge order, both by reducing long-range CDW correlations and by pinning CDWs to dopant sites[31]. Moreover, doping profoundly effects the material's band structure both by lifting the conduction band at the Gamma point and raising the Van Hove singularity towards the Fermi energy[17,32]. Disentangling these consequences of Sn-doping would be an interesting direction of future research, perhaps through a study that applies pressure to undoped samples, which affects the band structure similarly to electron doping, but does not induce disorder[32].

As noted, the mechanism responsible for the channelised supercurrents also appears to be influenced by the presence of a metallic Pt film deposited on the crystal. We observe a non-trivial redistribution of supercurrents near the Pt in both doped and undoped kagome devices. From our observations, we cannot determine the underlying origin of this behaviour. One possibility is that the deposited metallic film modifies the local charge order within the flake, impacting the CDW landscape in a manner similar to doping.

Next, we discuss the possible connection of the channelised supercurrents to reports of heterogeneous superconductivity in other materials. Heterogeneous superconductivity in a general sense has been previously observed in strongly-correlated systems such as the cuprates and iron-based superconductors[33,34]. Here, the superconducting gap modulates spatially, but superconductivity remains mostly finite over the entire system. The supercurrent distribution is therefore expected to be inhomogeneous throughout the sample, but not confined to spatially separated nanoscale channels. There are also reports on materials in which thin superconducting filaments and puddles coexist with large non-superconducting regions[35]. Examples include the superconducting LAO/STO interface[36], Ba(Fe,Co)As[37], and 1T-$TiSe_2$[38], where a competition between superconductivity and another order (e.g. charge or magnetic order) leads to filamentary superconductivity. In these cases, filamentarity is often associated with a percolating superconducting transition: upon cooling, an increasing fraction of the sample becomes superconducting, and the coupling between superconducting regions becomes progressively stronger[35]. Consequently, the fraction of the sample that can support supercurrents is expected to evolve strongly with temperature. This contrasts with our observations: the fraction of the sample that carries supercurrents does not expand with cooling, and the channelised supercurrents follow the same mesoscopic pattern over the entire temperature regime. The stability of this emergent flow pattern suggests that the observed channelisation of supercurrents does not arise from a conventional percolative transition.

If the channelised supercurrents indeed reflect a spatial separation of superconducting and non-superconducting regions, this raises an apparent contradiction: in previous studies on bulk crystals and powders, no signatures of fragmented superconductivity have been reported. In particular, for low Sn-doping, the superconducting fraction was found to be close to unity[17], underlining the absence of strong inhomogeneity. Phase separation can only emerge in the vicinity of critical points, where multiple orders compete. In bulk $CsV_3Sb_5$, superconductivity competes with charge density wave order. This competition can be affected in favour of superconductivity through, amongst others, pressure and stacking faults[39–41]. It is therefore natural to postulate that superconductivity is strengthened in regions where the charge density wave order is supressed, e.g. for structural reasons. Since the formation of the charge order is disfavoured by broken translation symmetry (in contrast to conventional superconductivity), an argument for enhanced superconductivity at boundaries and step edges of flake devices could be constructed.

It is important, however, to point out that in the undoped sample, where CDW domain walls are expected to be less abundant, supercurrent flows homogeneously throughout the interior of the flake. Nevertheless, we still observe more current accumulation near the edges than can be accounted for in London theory, indicating enhanced superconductivity at the sample boundaries. It remains an open question how the addition of doping, which introduces disorder and fragments the CDW, would so strongly channelise the supercurrents. It is also of crucial importance to understand how our nanoscale observations connect to bulk measurements of the diamagnetic response. We suggest that future research focuses on systematic studies of both thickness and doping dependence. Recent work has already hinted at thickness-dependent superconducting properties in undoped samples[42,43].

In conclusion, we have demonstrated a channelised supercurrent network in flakes of $CsV_3Sb_{5-x}Sn_x$ through observation of highly confined supercurrent paths and inhomogeneous Meissner screening. The channels are robust to thermal cycling up to room temperature and stable across and below the superconducting transition. They persist above the critical current of the flake, illustrating these critical currents are not related to depairing, but rather to Josephson coupling inside the channelised network. This surprising channelisation of supercurrent calls for a re-examination of the anomalous transport results that have been obtained in flake devices of $CsV_3Sb_5$ in the past, including those of higher-order Cooper pairing[1]. Beyond providing insight into the unique superconducting state of these flakes, our work stresses the importance of current imaging for characterisation of complex systems with competing orders.

## References

1. Ge, J. *et al.* Charge-4e and Charge-6e Flux Quantization and Higher Charge Superconductivity in Kagome Superconductor Ring Devices. *Phys. Rev. X* **14**, 021025 (2024).
2. Le, T. *et al.* Superconducting diode effect and interference patterns in kagome CsV3Sb5. *Nature* **630**, 64–69 (2024).
3. Le, T. *et al.* Thermomodulated Intrinsic Josephson Effect in Kagome CsV3Sb5. *Phys. Rev. Lett.* **135**, 096002 (2025).
4. Lou, H.-X. *et al.* Quantized radio-frequency rectification in a kagome superconductor Josephson diode. *Nat. Nanotechnol.* 1–7 (2026) doi:10.1038/s41565-025-02120-x.
5. Blom, T. J. *et al.* Emergent Network of Josephson Junctions in a Kagome Superconductor. *Nano Lett.* Article ASAP (2026) doi:10.1021/acs.nanolett.5c05587.
6. Zhou, S. & Wang, Z. Chern Fermi pocket, topological pair density wave, and charge-4e and charge-6e superconductivity in kagomé superconductors. *Nat. Commun.* **13**, 7288 (2022).
7. Lin, T.-Y., Song, F.-F. & Zhang, G.-M. Theory of the charge-6e condensed phase in kagome-lattice superconductors. *Phys. Rev. B* **111**, 054508 (2025).
8. Wang, Z., Zeng, K. & Wang, Z. Roton Superconductivity from Loop-Current Chern Metal on the Kagome Lattice. Preprint at https://doi.org/10.48550/arXiv.2504.02751 (2025).
9. Jiang, Y.-X. *et al.* Unconventional chiral charge order in kagome superconductor KV3Sb5. *Nat. Mater.* **20**, 1353–1357 (2021).
10. Guo, C. *et al.* Switchable chiral transport in charge-ordered kagome metal CsV3Sb5. *Nature* **611**, 461–466 (2022).
11. Xu, Y. *et al.* Three-state nematicity and magneto-optical Kerr effect in the charge density waves in kagome superconductors. *Nat. Phys.* **18**, 1470–1475 (2022).
12. Xing, Y. *et al.* Optical manipulation of the charge-density-wave state in RbV3Sb5. *Nature* **631**, 60–66 (2024).
13. Graham, J. N. *et al.* Depth-dependent study of time-reversal symmetry-breaking in the kagome superconductor AV3Sb5. *Nat. Commun.* **15**, 8978 (2024).
14. Elmers, H. J. *et al.* Chirality in the Kagome Metal CsV3Sb5. *Phys. Rev. Lett.* **134**, 096401 (2025).
15. Fernandes, R. M., Birol, T., Ye, M. & Vanderbilt, D. Loop-current order in kagome metals. *Nat. Phys.* 1–8 (2026) doi:10.1038/s41567-026-03229-z.
16. Wu, X. *et al.* Nature of Unconventional Pairing in the Kagome Superconductors AV3Sb5 (A = K, Rb, Cs). *Phys. Rev. Lett.* **127**, 177001 (2021).
17. Oey, Y. M. *et al.* Fermi level tuning and double-dome superconductivity in the kagome metal CsV3Sb5-xSnx. *Phys. Rev. Mater.* **6**, L041801 (2022).
18. Guguchia, Z. *et al.* Tunable unconventional kagome superconductivity in charge ordered RbV3Sb5 and KV3Sb5. *Nat. Commun.* **14**, 153 (2023).
19. Wilson, S. D. & Ortiz, B. R. AV3Sb5 kagome superconductors. *Nat. Rev. Mater.* **9**, 420–432 (2024).

20. Wang, S. *et al.* Switchable half-quantum flux states in a ring of the kagome superconductor CsV3Sb5. Preprint at https://doi.org/10.48550/arXiv.2512.10010 (2025).
21. Moler, K. A. Imaging quantum materials. *Nat. Mater.* **16**, 1049–1052 (2017).
22. Rog, M. *et al.* Tapping-Mode SQUID-on-Tip Microscopy with Proximity Josephson Junctions. *Nano Lett.* **26**, 1608–1615 (2026).
23. Kang, M. *et al.* Charge order landscape and competition with superconductivity in kagome metals. *Nat. Mater.* **22**, 186–193 (2023).
24. Pokharel, G. *et al.* Evolution of charge correlations in the hole-doped kagome superconductor CsV3-xTixSb5. *Phys. Rev. Mater.* **9**, 094805 (2025).
25. Xiao, Q. *et al.* Coexistence of multiple stacking charge density waves in kagome superconductor CsV3Sb5. *Phys. Rev. Res.* **5**, L012032 (2023).
26. Plumb, J. *et al.* Phase-separated charge order and twinning across length scales in CsV3Sb5. *Phys. Rev. Mater.* **8**, 093601 (2024).
27. Kautzsch, L. *et al.* Incommensurate charge-stripe correlations in the kagome superconductor CsV3Sb5−xSnx. *Npj Quantum Mater.* **8**, 37 (2023).
28. Rhoderick, E. H. & Wilson, E. M. Current Distribution in Thin Superconducting Films. *Nature* **194**, 1167–1168 (1962).
29. Ni, S. *et al.* Anisotropic Superconducting Properties of Kagome Metal CsV3Sb5. *Chin. Phys. Lett.* **38**, 057403 (2021).
30. Grant, M. J. *et al.* Superconducting energy gap structure of CsV3Sb5 from magnetic penetration depth measurements. *J. Phys. Condens. Matter* **37**, 065601 (2024).
31. Nikolov, I. K. *et al.* Observation of ubiquitous charge correlations and hidden quantum critical point in hole-doped kagome superconductors. Preprint at https://doi.org/10.48550/arXiv.2512.09256 (2025).
32. LaBollita, H. & Botana, A. S. Tuning the Van Hove singularities in AV3Sb5 (A = K, Rb, Cs) via pressure and doping. *Phys. Rev. B* **104**, 205129 (2021).
33. Pasupathy, A. N. *et al.* Electronic Origin of the Inhomogeneous Pairing Interaction in the High-Tc Superconductor Bi2Sr2CaCu2O8+δ. *Science* **320**, 196–201 (2008).
34. Cho, D., Bastiaans, K. M., Chatzopoulos, D., Gu, G. D. & Allan, M. P. A strongly inhomogeneous superfluid in an iron-based superconductor. *Nature* **571**, 541–545 (2019).
35. Venditti, G. *et al.* Superfluid response of two-dimensional filamentary superconductors. *SciPost Phys.* **15**, 239 (2023).
36. Biscaras, J. *et al.* Multiple quantum criticality in a two-dimensional superconductor. *Nat. Mater.* **12**, 542–548 (2013).
37. Xiao, H. *et al.* Filamentary superconductivity across the phase diagram of Ba(Fe,Co)2As2. *Phys. Rev. B* **86**, 064521 (2012).
38. Li, L. J. *et al.* Controlling many-body states by the electric-field effect in a two-dimensional material. *Nature* **529**, 185–189 (2016).
39. Chen, K. Y. *et al.* Double Superconducting Dome and Triple Enhancement of Tc in the Kagome Superconductor CsV3Sb5 under High Pressure. *Phys. Rev. Lett.* **126**, 247001 (2021).
40. Zhu, C. C. *et al.* Double-dome superconductivity under pressure in the V-based kagome metals AV3Sb5 (A = Rb and K). *Phys. Rev. B* **105**, 094507 (2022).
41. Han, X. *et al.* Atomic manipulation of the emergent quasi-2D superconductivity and pair density wave in a kagome metal. *Nat. Nanotechnol.* **20**, 1017–1025 (2025).
42. Song, B. *et al.* Anomalous enhancement of charge density wave in kagome superconductor CsV3Sb5 approaching the 2D limit. *Nat. Commun.* **14**, 2492 (2023).

43. Zhang, W. *et al.* Thickness-driven crossover from conventional to chiral nonreciprocal superconductivity in kagome metal CsV3Sb5. Preprint at https://doi.org/10.48550/arXiv.2605.29665 (2026).
44. Dynes, R. C., Narayanamurti, V. & Garno, J. P. Direct Measurement of Quasiparticle-Lifetime Broadening in a Strong-Coupled Superconductor. *Phys. Rev. Lett.* **41**, 1509–1512 (1978).

# Methods

## Crystal synthesis

Single crystals of $CsV_3Sb_{5-x}Sn_x$, x=0 and x=0.025 were synthesized by the flux growth method. Elemental Cs (Alfa 99.98 %), V powder (Sigma 99.9 %), Sb shot (Alfa 99.999%), and Sn shot (99.999 %) were weighed out in 20:15:120:0, and 20:15:117.5:10.9 ratios respectively in an inert argon environment ($H_2O$ and $O_2$ < 0.5 ppm) . The starting materials were loaded into a tungsten carbide vial and milled in a SPEX8000D mixer for 60 min. The milled powders were packed in alumina crucibles and sealed inside a steel tube for growth. The tubes were heated up to 1000 °C, held at this temperature for 12 hours, cooled down to 950 °C at 5 °C/hour and further cooled at 1 °C/hour to 500 °C. The resulting hexagonal crystals were manually extracted at room temperature in air. Their effective doping was analysed with a scanning electron microscope Hitachi TM4000Plus with energy dispersive spectroscopy. 2H-$NbSe_2$ crystals were purchased from HQ Graphene.

## Device fabrication

All devices are fabricated by first exfoliating flakes from the parent crystal using the scotch tape method. After optically selecting good flakes, we use electron beam lithography followed by RF-sputtering to deposit four Pt contacts on each flake in a standard lift-off process. For sample preservation, the samples are sputter coated with silicon oxide and stored in vacuum when not in use. The thickness and structure of the flake is measured using high-resolution room temperature atomic force microscopy (JPK NanoWizard 4).

Before SQUID-on-tip imaging, transport experiments are conducted on all flakes. Resistance-temperature and superconducting interference experiments were performed inside a continuous flow cryostat using a Zurich Instruments MFLI lock-in amplifier.

## Tapping-mode SQUID-on-tip Microscopy

We use Tapping-Mode SQUID-on-tip microscopy as previously described in ref.[22]. To prepare the probes, proximity junction (Nb-Cu-Nb) SQUIDs are created at the apex of a commercial tuning fork atomic force microscopy (TF-AFM) probe using Ga focused ion beam milling. In this work, we use two probes of effective SQUID diameters 260 and 290 nm.

The sample and probe are loaded together inside a cryostat that is equipped with a Tapping-Mode SQUID-on-tip microscope. Devices D1 and D2 were measured inside a continuous flow cryostat (Oxford Instruments) with a base temperature of 1.7 K and equipped with a vectorial superconducting magnet, whereas Devices D3 and D4 were measured inside a dry cryostat (Bluefors) with a base temperature of 900 mK and equipped with a home-made single-axis in-plane superconducting magnet. All sample lines are cryogenically filtered up to 10 GHz. To optimize SQUID performance, all measurements on device D1 and D2 are conducted in a small in-plane field of 5-8 mT. To prevent the nucleation of vortices in devices D3 and D4, these measurements are conducted at zero external field, unless specified otherwise.

All images are acquired in TF-AFM mode. In this mode, the tip-sample separation is automatically minimized, and is limited only by the oscillation amplitude of the cantilever.

To measure Biot-Savart fields generated by transport currents, we apply a.c. currents at a frequency of 283 Hz. Generally, we apply this current as a sine wave. An exception is the current-dependent measurements of Fig. 3c, where the application of a square wave allows us to measure only the current distribution at a single point on the current-voltage characteristic. Biot-Savart fields are then resolved by demodulating the SQUID response at the applied frequency. To measure static fields, we use the gradiometric technique described in ref.[22] by oscillating the sample 35 nm at a frequency of 327 Hz.

## Scanning Tunnelling Microscopy Characterisation

Scanning tunnelling microscopy (STM) measurements on the bulk $CsV_3Sb_{5-x}Sn_x$ crystal with x=0.03 were performed using a superconducting tip. For these STM measurements, we used a modified commercial scanning tunnelling microscope (USM-1500, Unisoku Co., Ltd) and a mechanically sharpened Pt-Ir wire for the STM tip, which was coated with Pb by indenting it into a bulk Pb crystal. The crystal was cleaved in situ in an ultrahigh vacuum ($P \approx 10 \times 10^{-10}$ mbar) at low temperature (T≈30 K) and were transferred to the pre-cooled (T≈2.2 K) STM head as quickly as possible to minimize surface contamination.

The differential conductance spectrum exhibits a finite superconducting gap determined by two coherence peaks. To fit the data, we use a modified Dynes formula for the density of states of the tip and sample, $D_{\mathrm{t(s)}}$[34,44]:

$$D_{\mathrm{t(s)}} = \mathrm{Re}\left[\mathrm{sgn}(\epsilon)\frac{\epsilon}{\sqrt{\epsilon^2 + 2i\gamma\epsilon - \Delta^2_{\mathrm{CP,t(s)}}}}\right]$$

Here, $\epsilon$ is energy, $\gamma$ is a phenomenological broadening term, and $\Delta_{\mathrm{CP}}$ is the gap between the coherence peaks. We measured multiple spectra at different positions, and confirmed the consistency of the superconducting gap within the error.

## Author Information

### Corresponding Author

Kaveh Lahabi, Huygens-Kamerlingh Onnes Laboratory, Leiden University, P.O. Box 9504, 2300 RA Leiden, The Netherlands; Email: lahabi@physics.leidenuniv.nl

## Acknowledgement

We thank J. Aarts, S. Sanna, G. Venditti and L. Rademaker for insightful discussions, and M.H. Hesselberth, S. van Leeuwen, L.J. Wigbout and F. Galli for their technical assistance. This work was financed by the Dutch National Growth Fund (NGF), as part of the Quantum Delta NL Programme in the Project SME_R027_MQC, and the Dutch Research Council, as part of the “The Full Picture: Novel Microscopy with Smart Quantum Probes” (Project: VI.Veni.212.302),Take-off Phase 2 (Project: 20964), and Quantum Limits (Project: SUMMIT.1.1016). It was also partly supported by European Union via the NextGenerationEU program. S.D.W. acknowledges support from the U.S. Department of Energy (DOE), Office of Basic Energy Sciences, Division of Materials Sciences and Engineering under grant no.DE-SC0020305. A.C.S. acknowledges support via the UC Santa Barbara NSF Quantum Foundry funded via the Q-AMASE-i program under award DMR-1906325. J.L. acknowledges support from the European Research Council (ERC Advanced Grant No. 101095574). T.N. acknowledges support from Swiss National Science Foundation through a Consolidator Grant (iTQC, TMCG-2_213805). AFM measurements were performed in the AFM Facility of the Leiden Institute of Physics (LION).

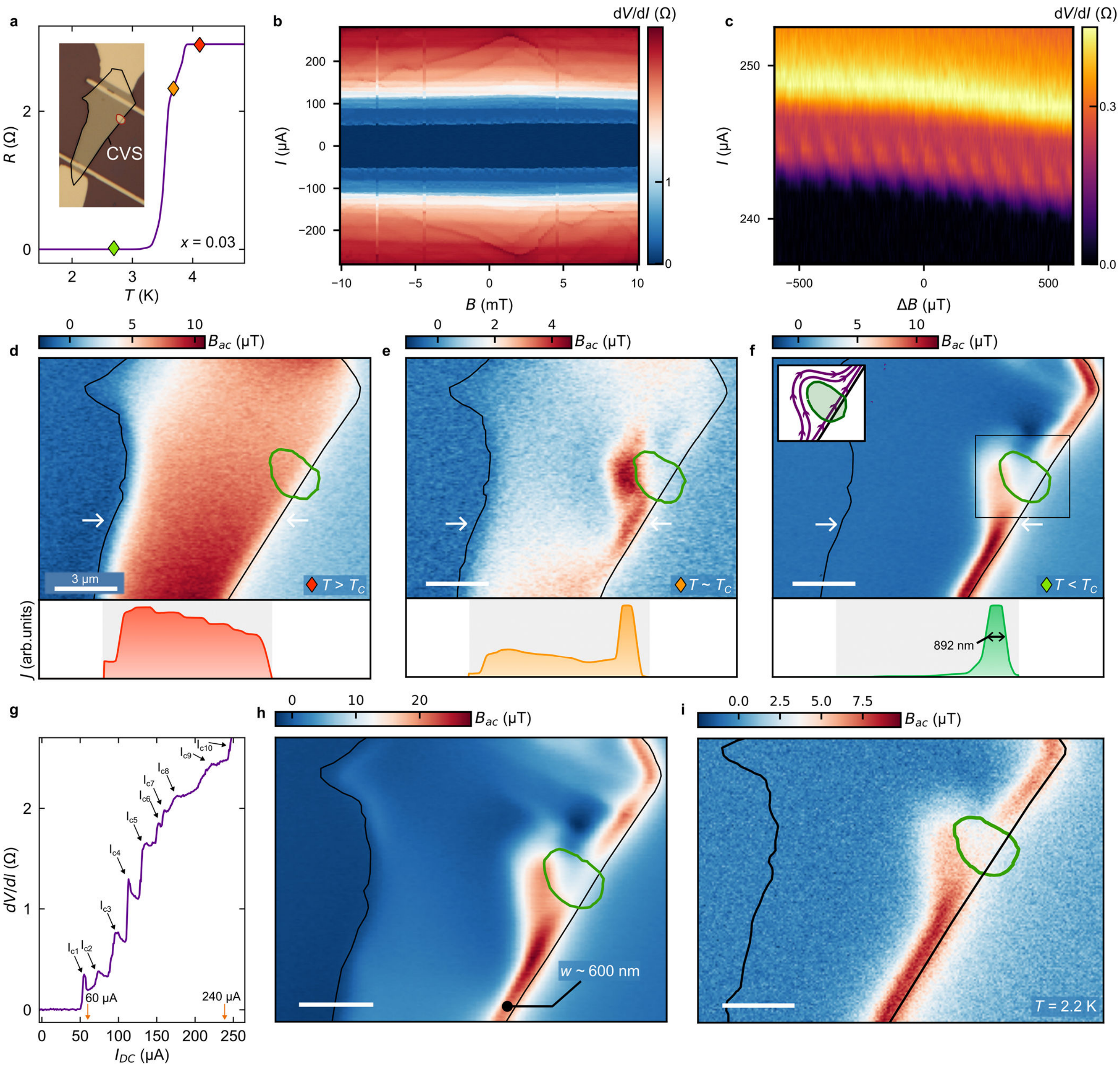

**Extended Data Fig. 1 | Overview of our measurements on Device D2 ($x$ = 0.03). a**, Resistance-temperature characteristic of the superconducting transition, showing broadening. Inset: optical micrograph of the flake, with the Pt deposit highlighted in red. **b**,**c**, supercurrent interference patterns for various field scales taken at 2.8 K (**b**) and 1.5 K (**c**), showing oscillations for many different interference areas manifest inside this flake. **d-f**, TM-SOT images of current flow (peak amplitude 240 μA in (d), 60 μA in (e-f)) above, at, and below Tc, showing the emergence of a single stable channel bound to the right edge of the flake. At the Pt deposit, the current flow is impacted, and it bends around the island (cartoon in inset). Bottom panels show current reconstruction along the line indicated by the arrows. The full-width at half-maximum (FWHM) of the channel, measured normal to the flake boundary, is 892 nm. **g**, Current-voltage characteristic of the sample showing a cascade of critical currents without ever reverting to the normal-state resistance. **h**, High-current image, taken with an amplitude of 240 μA ($I > I_{c10}$). The channel stays completely fixed and does not decay, reaffirming that none of these critical currents result in depairing. At these large currents, other parts of the network become visible. As indicated, the FWHM of the channel away from the Pt deposit is a constant 600 nm. **i**, Image of current flow at a lower temperature (2.2 K) taken after a thermal cycle to room temperature, showing that the channelised configuration is stable below $T_c$ and is not changed upon thermal cycling.

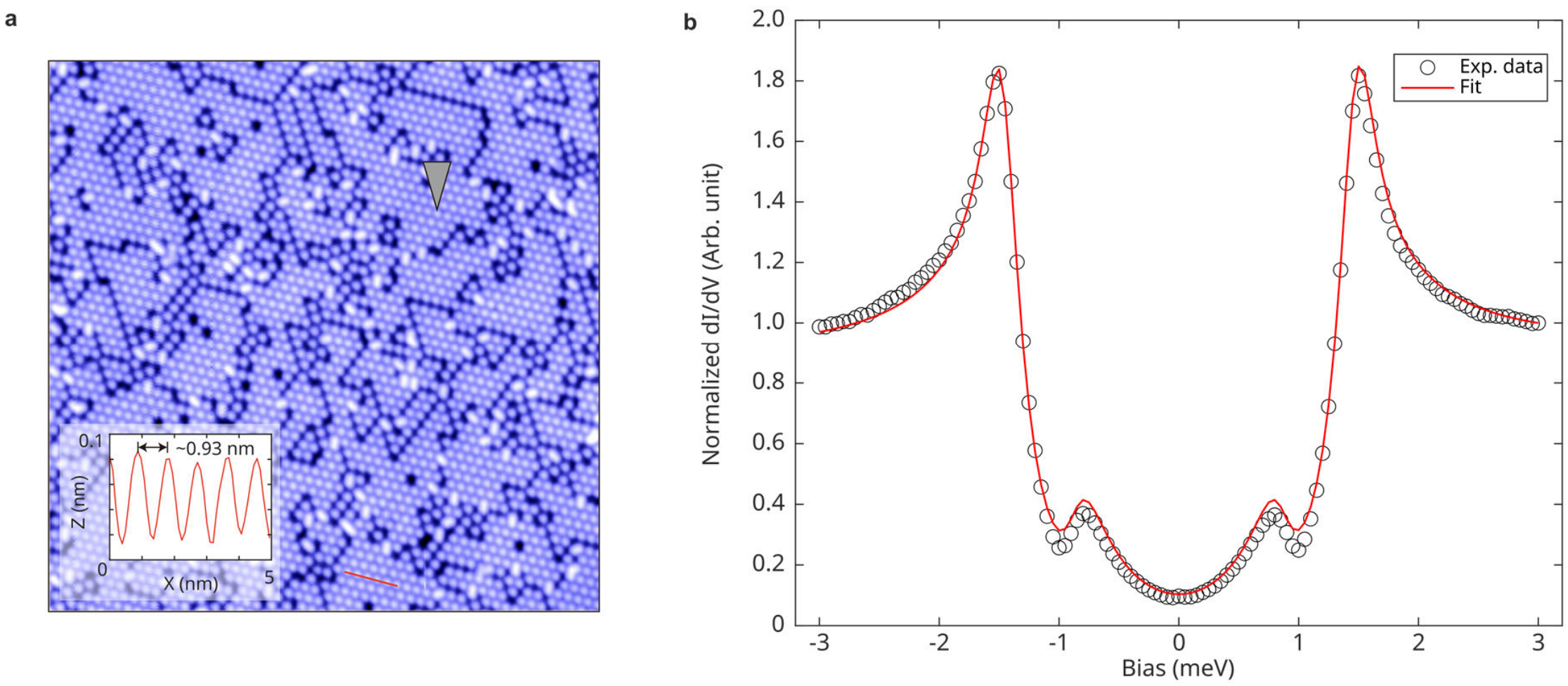

**Extended Data Fig. 2 | Scanning tunnelling microscopy study on superconductivity in the parent x = 0.03 crystal. a**, STM topography of the Cs-terminated surface (50 nm x 50 nm, $V_{\text{bias}}$ = 200 mV, $I_{\text{set}}$ = 100 pA), showing the $\sqrt{3} \times \sqrt{3}$ structure (line profile shown in the inset). **b**, dI/dV spectrum (open circles) measured with a superconducting tip. The measurement location is indicated in the topography in **a**. We obtain a superconducting gap of 0.30±0.01 meV by fitting the data using a modified Dynes formula, as described in the Methods.

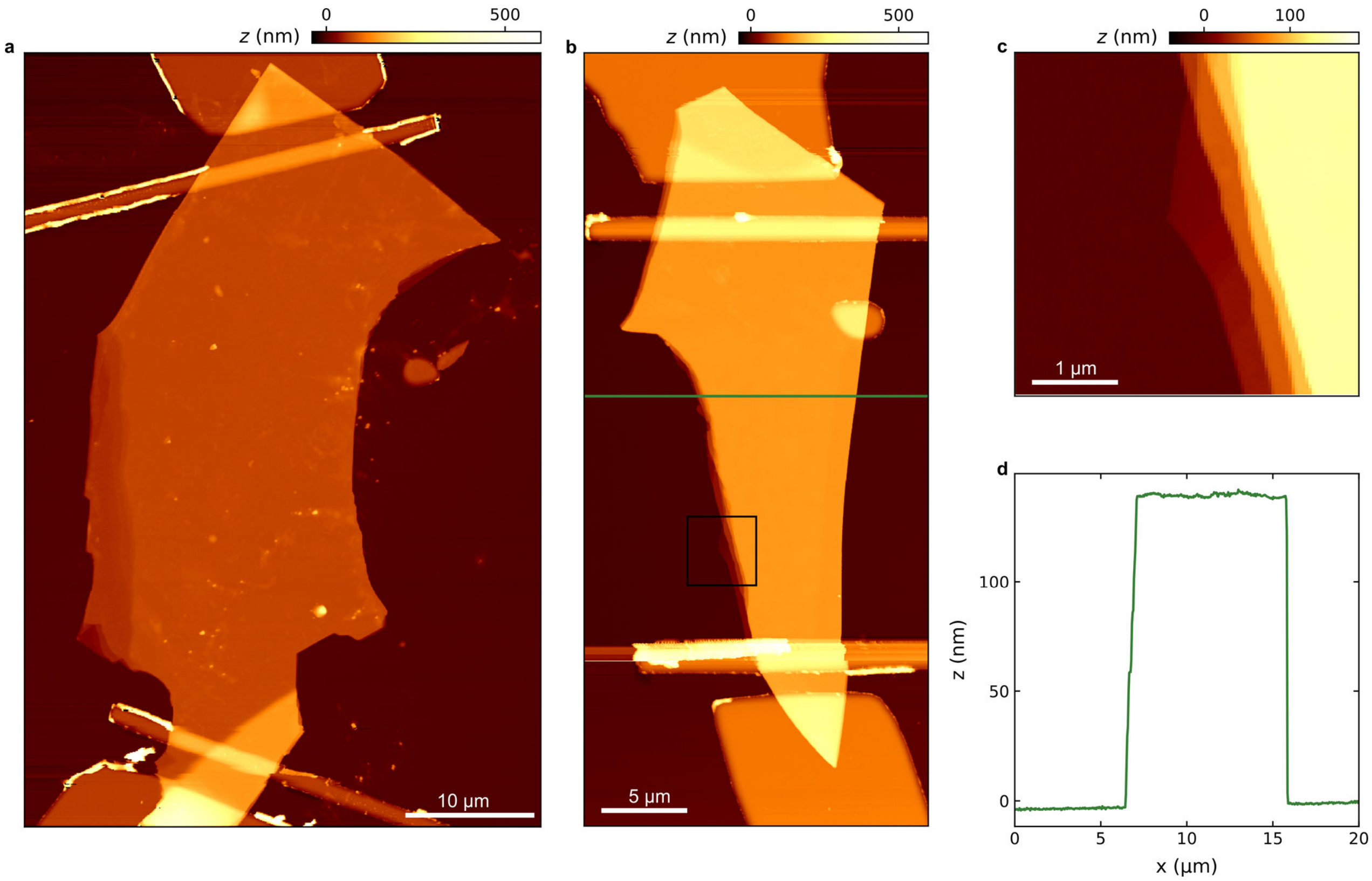

**Extended Data Fig. 3 | High-resolution atomic force microscopy on the doped devices D1 and D2. a**, AFM micrograph of device D1. Apart from the terracing on the left edge, the device is flat. Streaks and blotches in the image are tape residue from the exfoliation process. **b**, AFM micrograph of device D2. While the left edge is strongly terraced, the right edge is sharp. The top of the flake is flat. **c**, Zoom in of **b** in the region marked in black, showing the terracing of the left edge. **d**, Line-cut of **b** along the green line, displaying that the top of the flake is flat.

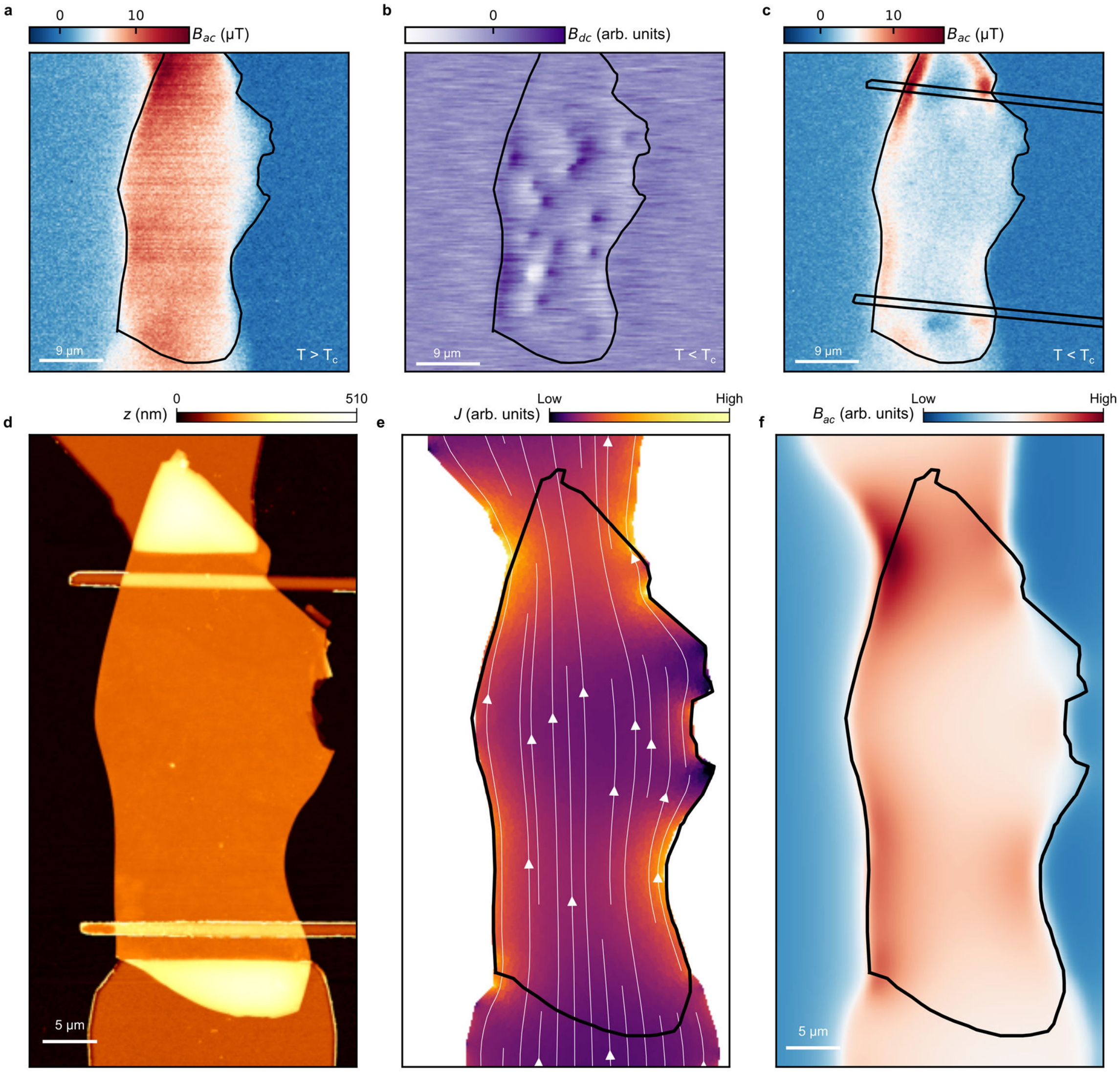

**Extended Data Fig. 4 | Extended measurements on device D3 (undoped). a**, TM-SOT image of current flow above the critical temperature ($T = 6$ K). **b**, TM-SOT image of the DC field above the flake below $T_c$ (T = 1 K) showing the uniform nucleation of vortices at low fields. **c**, TM-SOT image of current flow in the flake below $T_c$ (T = 1K) below the critical current. Same image as main text Fig. 2f., added for comparison. Here, we additionally mark the position of the platinum leads, emphasizing that the current distribution is especially inhomogeneous near the metallic contacts. **d**, Room-temperature AFM image of the device, showing a completely flat top layer. **e**, Simulated current flow through device D3 in the superconducting state. From the AFM imaging, we extract the flake and contact geometry. Simplifying the flake-contact system to a single superconductor of uniform thickness (t = 118 nm) and penetration depth ($\lambda$ = 500 nm), we can calculate the current flow path using the London equations. Details of the simulations are provided in the Supplementary Information. **f**, Simulated TM-SOT image corresponding to the currents in **(e)**, using the properties of the probe in **a-c**. While the same qualitative features are visible, we see that the supercurrents in **c** are much more confined than expected in London theory, especially near the electrical contacts, implying that in these undoped flakes a substantial current inhomogeneity is already present.

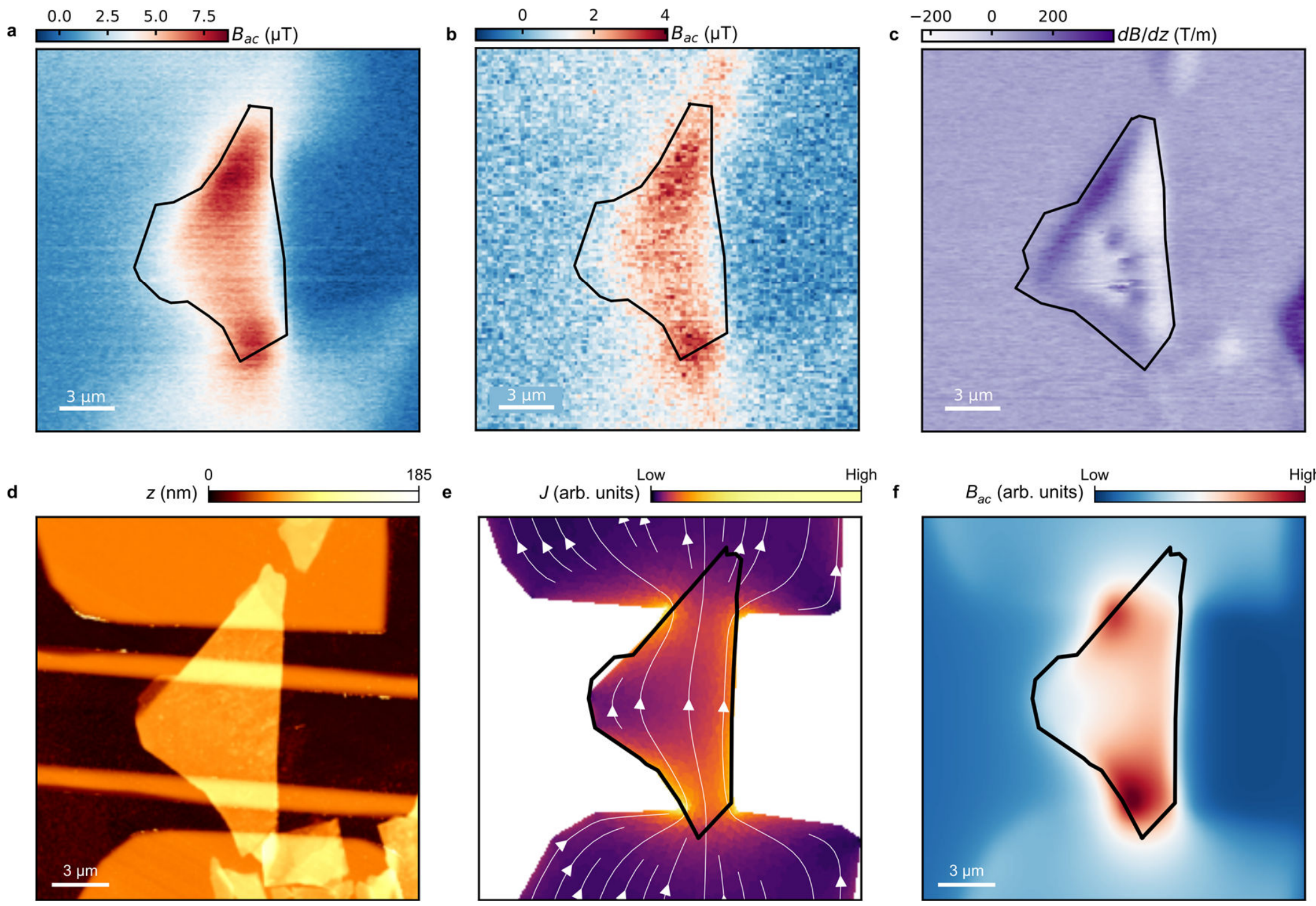

**Extended Data Fig. 5 | Extended measurements on device D4 (2H-$NbSe_2$). a**, TM-SOT image of current flow above the critical temperature ($T$ = 7.5 K). **b**, TM-SOT image of current flow close under $T_c$ ($T$ = 6.5 K) showing a uniform current profile immediately emerges. **c**, TM-SOT gradiometric image of the field gradients above the sample at $T$ = 1 K. Meissner currents are clearly visible at the edge of both the contacted flake and the uncontacted flake fragments that lie in the field of view. In the middle of the flake, three vortices are visible. We attribute the line noise around the vortices to the lack of pinning in this single-crystal flake. **d**, Room-temperature AFM image of the device, showing a completely flat top layer. **e**, Simulated current flow through device D4 in the superconducting state. From the AFM imaging, we extract the flake and contact geometry. Simplifying the flake-contact system to a single superconductor of uniform thickness (t = 51 nm) and penetration depth (λ = 130 nm), we calculate the current flow path using the London equations. Details of the simulations are provided in the Supplementary Information. **f**, Simulated TM-SOT image corresponding to the currents in **e**, using the properties of the probe in **a-c**. The same qualitative features are visible. The theory predicts more current crowding at the flake-contact corners, but this can be fully attributed to the limitations of our approximation that the contacts themselves are part of the superconductor.

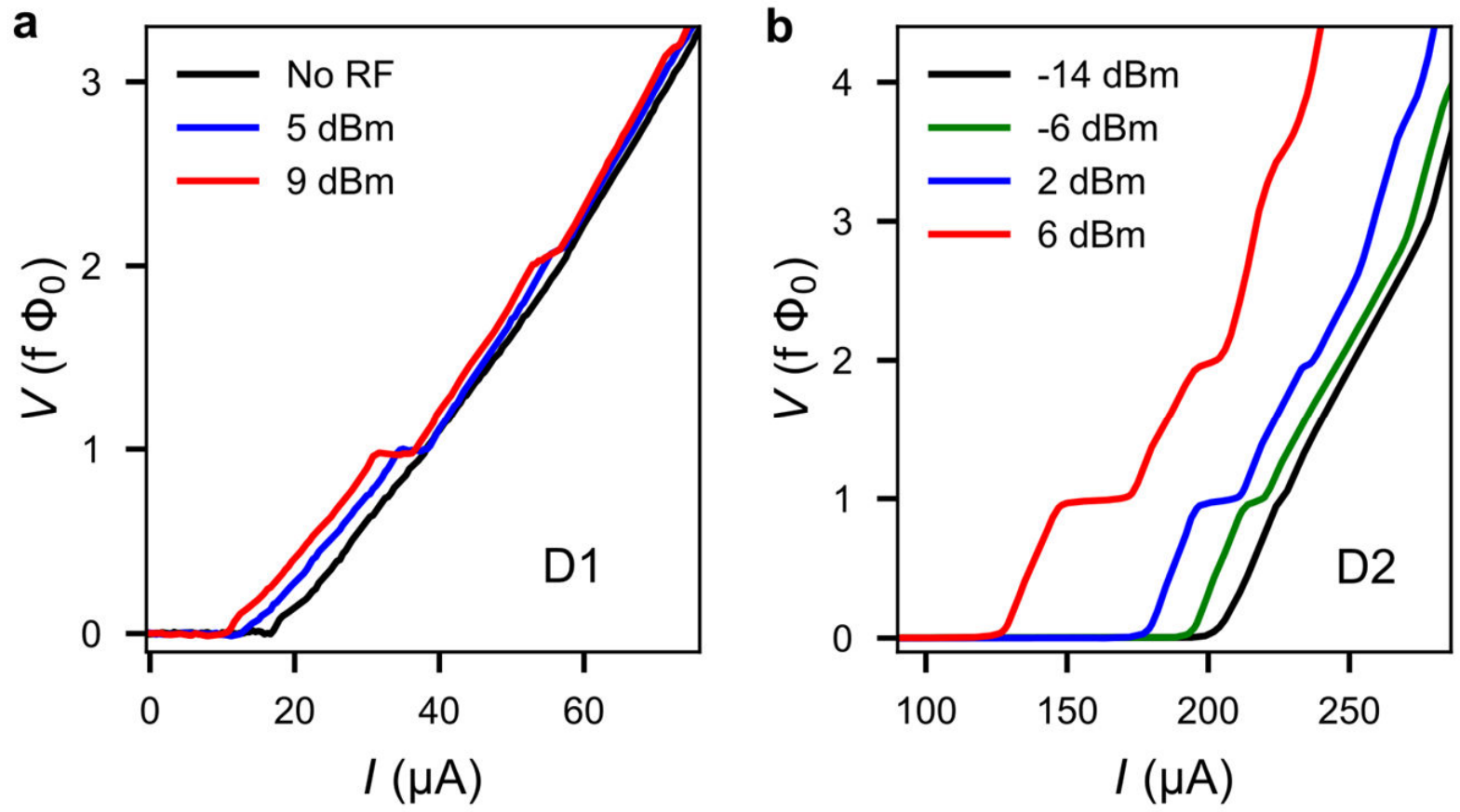

**Extended Data Fig. 6 | Shapiro response of both Sn-doped devices. a**, Current-voltage characteristics of device D1 with, and without the application of external 4 GHz radiation of varying power. **b**, Current-voltage characteristics of device D2 with 3370 MHz radiation. Both sets of measurements were performed at 1.4 K, well below the critical temperature. In both cases, voltage is shown in units of frequency times the flux quantum (f $\Phi_0$). No Shapiro steps were observed in devices D3 and D4 when applying radio-frequency radiation.

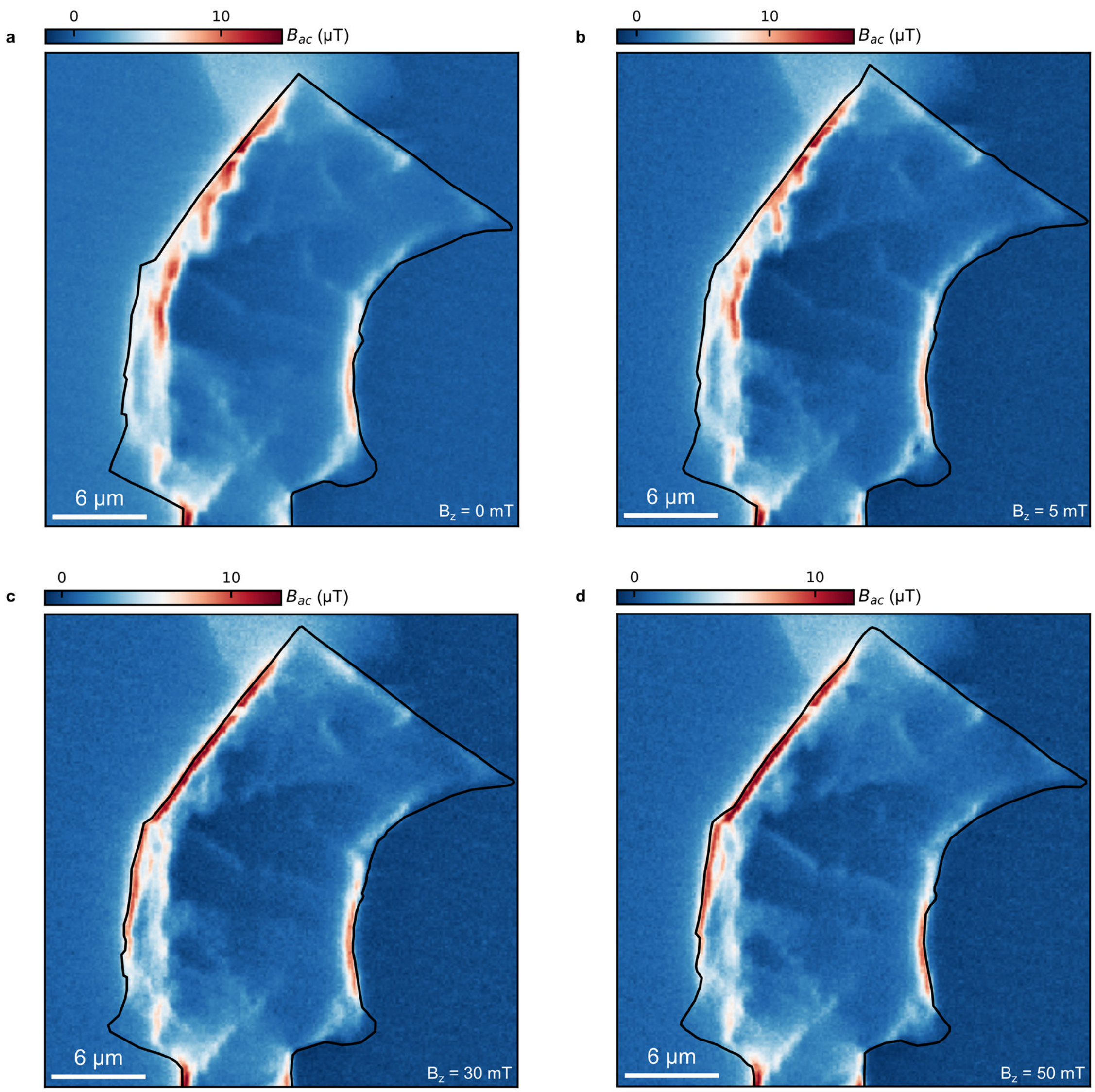

**Extended Data Fig. 7 | Current distribution at various out-of-plane magnetic fields in device D1. a-d**, TM-SOT images taken at 0, 5, 30 and 50 mT out-of-plane applied fields, with an applied current of 85 μA. The out-of-plane current does not break any channels, nor does it seem to move them. However, the distribution of currents throughout the network changes considerably, especially at the north-western edge. In a network of Josephson junctions, the current distribution is naturally changed by an applied field, since its flux induces an additional phase around each network cell.

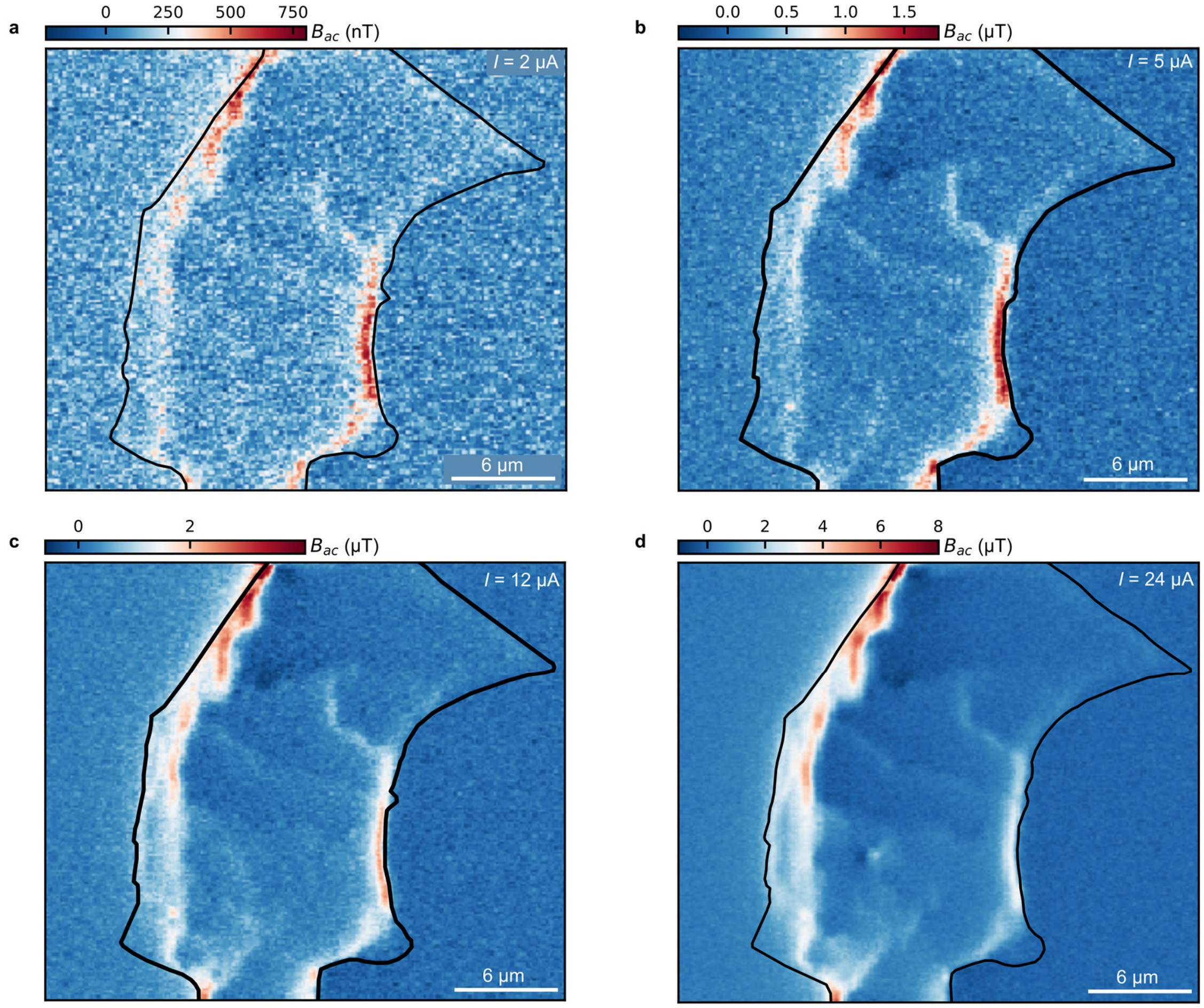

**Extended Data Fig. 8 | Current distribution at various injection currents in device D1. a-d**, TM-SOT images taken at 2, 5, 12 and 24 μA applied injection current. Current is applied as a square wave to lock in on one particular current distribution. As emphasized in Fig. 3c and 3f of the main text, the current distribution is strongly dependent on applied current. As the current is raised through the first critical currents, the right channel loses prominence and the current becomes more equally distributed over the left and right channels.